\documentstyle[preprint,aps]{revtex}
\begin{document}

\draft
\begin{title}
{Maximum Overheating and Partial Wetting of Nonmelting Solid Surfaces}
\end{title}

\author{Francesco D. Di Tolla, Furio Ercolessi, and Erio Tosatti}
\address{
International School for Advanced Studies (SISSA-ISAS),
Via Beirut 4, I-34014 Trieste, Italy
}

\maketitle

\begin{abstract}
Surfaces which do not exhibit surface melting below the melting point
(nonmelting surfaces) have been recently observed to sustain a very large
amount of overheating.  We present a theory which identifies a maximum
overheating temperature, and relates it to other thermodynamical properties
of the surface, in particular to geometrical properties more readily accessible
to experiment.  These are the angle of partial wetting, and the
nonmelting-induced faceting angle.  We also present molecular dynamics
simulations of a liquid droplet deposited on Al(111), showing lack of spreading
and partial wetting in good agreement with the theory.
\end{abstract}



\narrowtext

For a long time crystal overheating above the bulk melting temperature
$T_m$ has been believed to be impossible, at least
in the presence of a free clean surface.
The standard argument \cite{frenkel,dash} is that
surface premelting will always take place and act as an ubiquitous seed
for the liquid to grow.
The well-known surface melting of Pb(110) \cite{vanderveen,pluis}
provided a first microscopic evidence of how liquid nucleation
takes place on a solid below $T_m$.
It was only a little later that simulations of Au(111) \cite{carnevali}
and newer experiments on Pb(111) \cite{frenken} and Al(111) \cite {gonAl}
demonstrated microscopically that the opposite could also happen, namely that
certain surfaces may exhibit nonmelting up to and in fact even
{\em above} the melting point \cite{carnevali,tos1}.
A solid bounded by such surfaces can therefore be {\em overheated},
although in a metastable state, above $T_m$.
M\'etois {\em et al.} have first shown that small Pb particles
with strictly (111) facets are easily overheated by a few degrees above $T_m$
\cite{metois}.
Even more strikingly, Herman {\em et al.} found that a flat nonmelting Pb
surface can be overheated by as much as 120 K above $T_m$ \cite{herman}.
This implies that the free energy of a crystal surface can have
a local minimum for zero liquid thickness. As in other nucleation problems
one should thus expect the metastable overheated state to survive up to
some instability temperature $T_i > T_m$, where the barrier
finally disappears (fig.\ \ref{fig:df}, inset).
At present, however, there is no further available understanding
of this phenomenon. In particular, there are no means
to calculate $T_i$ and possibly connect it with other quantities
which are more readily measurable in a surface experiment.
At a more microscopic level, it is very desirable to understand the different
behavior of a nonmelting and of a melting surface, against nucleation of
the liquid.

In this Letter, we introduce a simple theory of
surface nonmelting which predicts the existence of a $T_i$,
and connects its value with apparently unrelated geometrical quantities.
These are the partial wetting angle $\theta_m$ which a drop of melt
will form with that crystal surface at $T=T_m$,
and the faceting angle $\theta_c$ of a vicinal surface.
The angle $\theta_m$ has also been rather commonly measured in the past,
a few early examples being the (0001) face of Cd \cite{mut1}
and the (100) faces of several alkali halides \cite{mut2}.
The nonmelting-induced faceting \cite{bila,helene} angle $\theta_c$ has been
well characterized experimentally and theoretically for (111) vicinals of
Au \cite{bila,stockAu}, Cu \cite{stockCu} and
Pb \cite{bila,helene,pavlovska,heyraud}.
The connection we find between $\theta_m$, $\theta_c$ and $T_i$ offers
new insight in nonmelting surfaces. At a microscopic level,
we substantiate this connection with molecular dynamics (MD)
simulations of Al(111), which demonstrate both the non-spreading
of a liquid drop at $T_m$, and the overheating of the flat face.
The predicted relationship between $\theta_m$, $\theta_c$ and $T_i$
is found to be in excellent agreement with the simulation results,
as well as with experiments.

{\em (i) Theory:}
Consider a liquid film of thickness $\ell$, sandwiched between semiinfinite
solid and vapor, and let $\ell$ grow from zero (no liquid) to a finite value.
The change in free energy per unit area takes the standard form \cite{pluis}
\begin{equation}
\Delta F(\ell)=\rho L \ell (1-T/T_m) + \Delta \gamma (\ell)
\label{eq:oh}
\end{equation}
where $\rho$ is the liquid density, $L$ the latent heat of melting, and
$\Delta \gamma (\ell)$
the difference between the overall free energy of the two interacting
solid-liquid (SL) and liquid-vapor (LV) interfaces
separated by a distance $\ell$,
and the free energy of the solid-vapor (SV) interface.
By definition, $\Delta\gamma (0) = 0$.
Assuming short-range forces only, this term can phenomenologically be written
as $\Delta\gamma(\ell)=\Delta\gamma_{\infty}[1-\exp(-\ell/\xi)]$
where
$\Delta \gamma_{\infty}\equiv \gamma_{\rm SL}+\gamma_{\rm LV}-\gamma_{\rm SV}$
is the net free energy change upon conversion of the SV
interface in two non-interacting SL and LV interfaces, and $\xi$ is
a correlation length in the liquid.
For a melting surface $\Delta \gamma_{\infty}<0$, and,
for $T_w < T < T_m$, $\Delta F$
will have a minimum at
$\ell_0(T) = \xi\ln\left[T_m|\Delta\gamma_{\infty}|/(T_m-T) L \rho \xi\right]$
which  is the mean-field thickness of the melted film \cite{pluis}.
The wetting temperature defined by $\ell_0(T_w)=0$ is
$T_w=T_m\left(1-|\Delta\gamma_{\infty}|/L \rho  \xi \right)$.

For a {\em nonmelting surface}, $\Delta\gamma_{\infty}>0$, and we
move over to $T>T_m$.
Here, $\Delta F(\ell)$ will instead have a local minimum at
$\ell=0$, the absolute minimum for
$\ell\rightarrow\infty$, and a {\em maximum} at a critical thickness
\begin{equation}
\ell_c(T) = \xi\ln\left[\frac{T_m \Delta\gamma_{\infty} }
{(T-T_m) L \rho \xi}\right]
\label{eq:ellc}
\end{equation}
as shown in fig.\ \ref{fig:df}.
The local minimum at $\ell=0$ signifies metastability of the crystalline
surface for $T<T_i$, the maximum overheating temperature. The minimum
disappears when $\ell_c(T_i)=0$, yielding
\begin{equation}
T_i=T_m\left(1+\frac{\Delta\gamma_{\infty}}{L \rho  \xi}\right) .
\label{eq:ti}
\end{equation}
Above $T_i$, the crystal surface will melt,
no matter what its initial state is.
In particular a surface which is initially crystalline will wet itself with a
liquid film, which will grow, and gradually melt the whole crystal.
Hence $T_i$ can be seen as a non-equilibrium wetting temperature, or, more
accurately, as a {\em spinodal} point for the overheated solid surface.
For $T_m < T < T_i$, the predicted
behavior is that typical of a nucleation problem.
If the surface is prepared initially with a
melted film of thickness $\ell>\ell_c$ (upper vertical arrow in
fig.\ \ref{fig:df}),
then melting will proceed, and $\ell$ will grow to infinity, reaching full
equilibrium. If, conversely, the starting thickness is less than $\ell_c$,
then the surface will recrystallize, to reach the local, metastable
minimum at $\ell=0$ (lower arrow in fig.\ \ref{fig:df}).
This peculiar behavior was first found and described in detail in an early
simulation of the nonmelting surface Au(111) \cite{carnevali}.

We now show that there is a simple connection
between $T_i$, and the macroscopic non-wetting angle $\theta_m$ at $T=T_m$.
Following Nozi\'eres \cite{nozi}, the angles
$\theta_{\rm LV}$, $\theta_{\rm SL}$,
formed by a drop of melt onto a nonwetting surface of
the same material (fig.\ \ref{fig:drop}), satisfy the equations
\begin{equation}
\gamma_{\rm SV} = \gamma_{\rm LV} \cos \theta_{\rm LV}
                + \gamma_{\rm SL} \cos \theta_{\rm SL}
\label{eq:gamma}
\end{equation}
\begin{equation}
R_{\rm LV} \sin \theta_{\rm LV} = R_{\rm SL} \sin \theta_{\rm SL}
\label{eq:R}
\end{equation}
where $R_{\rm LV}$, $R_{\rm SL}$ are the radii of respectively the
LV and SL drop boundaries (supposedly spherical).
Eq.\ (\ref{eq:gamma}) is simply the balance of lateral forces, while
eq.\ (\ref{eq:R}) follows from simple geometry.
Laplace's pressure equation $P=2\gamma / R$ determines the shape ratio
$x(T)\equiv R_{\rm LV}/R_{\rm SL} = \sin\theta_{\rm SL}/\sin\theta_{\rm LV} =
[\gamma_{\rm LV} P_{\rm SL}(T)]/[\gamma_{\rm SL} P_{\rm LV}(T)]$.
Since $P_{\rm SL} \propto (T-T_m)$ near $T_m$,
we expect $\theta_{\rm SL}$ to switch
from negative for $T<T_m$ to positive for $T>T_m$. At $T=T_m$,
$x=\theta_{\rm SL}=0$, $R_{\rm LV}= \infty$, and
$\theta_{\rm LV}=\theta_m$ where
\begin{equation}
\cos\theta_m = 1 - \frac{\Delta\gamma_\infty}{\gamma_{\rm LV}} .
\label{eq:thetam}
\end{equation}
Comparison of (\ref{eq:thetam}) with (\ref{eq:ti}) shows that knowledge of
$\Delta\gamma_\infty$ at $T=T_m$ determines {\em both} $T_i$ and
$\theta_m$, which are monotonically related by
\begin{equation}
\frac{T_i}{T_m} =
     1 + \frac{2\gamma_{\rm LV}}{L\rho\xi} \sin^2\frac{\theta_m}{2} .
\label{eq:tithetam}
\end{equation}

For a nonmelting surface there is a second important angle $\theta_c$,
which is the nonmelting-induced faceting angle \cite{bila,helene}.
Consider vicinal faces tilted at an angle $\theta$ away from
the nonmelting face.
At $T=T_m$ there are two well-defined free energy minima (solid, $\ell=0$
and liquid, $\ell=\infty$). We can thus draw \cite{nozi}
the two projected surface free energy branches
$\sigma(\theta)=\gamma(\theta)/\cos\theta$ as a function of
the step density $t=|\tan\theta|$.
The two branches are approximately
given by the standard expressions
\begin{equation}
\sigma_S(\theta)=\gamma_{\rm SV}+\mu t + g t^3
\label{eq:sigs}
\end{equation}
\begin{equation}
\sigma_L(\theta)=\left(\gamma_{\rm LV}+\gamma_{\rm SL}\right) \sqrt{1+t^2}
\label{eq:sigl}
\end{equation}
where $\mu$ and $g$ are the step free energy and the step-step repulsion on the
solid surface. Here we have further assumed that $\gamma_{\rm SL}$ is
approximately independent of $\theta$.
The faceting angle is
given by $\theta_c=\arctan t_c$ which satisfies the
double tangent construction:
$c_0+c_1t_0=\gamma_{\rm SV}+\mu t_0 + g t_0^3$,
$c_1=\mu+3gt_0^2$, and
$c_0+c_1t_c=\left(\gamma_{\rm LV}+\gamma_{\rm SL}\right) \sqrt{1+t_c^2}$,
$c_1={(\gamma_{\rm LV}+\gamma_{\rm SL})\,t_c}/{\sqrt{1+t_c^2}}$.
A particularly simple solution is obtained if the cubic (step-step repulsion)
term $gt_0^3$ can be ignored, whence $t_0=0$, $c_0=\gamma_{\rm SV}$,
$c_1=[(\gamma_{\rm LV}+\gamma_{\rm SL})^2 - \gamma_{\rm SV}^2]^{1/2}$
(note the nonanalyticity of $\sigma_S$ at $t=0$) and
$t_c=\{ [\left(\gamma_{\rm LV}+\gamma_{\rm SL}\right)/\gamma_{\rm SV}]^2
        -1\}^{1/2}$.
Even when this approximation cannot be made, and $t_0$ is nonzero (as is
the case for Pb(111) \cite{helene}, where $\arctan t_0\simeq 2^\circ$),
the above is still a pretty good approximation to the faceting angle,
which is therefore simply related to $\Delta\gamma_\infty$:
\begin{equation}
\cos\theta_c\simeq
\left(1+\frac{\Delta\gamma_\infty}{\gamma_{\rm SV}}\right)^{-1} .
\label{eq:thetac}
\end{equation}
Eq.\ (\ref{eq:gamma}) shows that $\theta_c$ is identical to {\em both} the
droplet angles $\theta_{\rm LV}$, $\theta_{\rm SL}$ at a single temperature
$T_u>T_m$, satisfying $x(T_u)=R_{\rm LV}/R_{\rm SL}=1$.
We note that $\theta_c$ is slightly smaller than
$\theta_m$. The outer droplet angle $\theta_{\rm LV}$
will therefore decrease from $\theta_m$ to $\theta_c$ to zero when $T$ is
raised from $T_m$ to $T_u$ to $T_i$.
Finally, we observe that a physical upper bound for $\theta_m$ and $\theta_c$
is given by  $\Delta \gamma_\infty \ll \gamma_{\rm LV}$, whence
$\theta_m\ll90^\circ$, and  $\theta_c\ll60^\circ$,
i.e., a melt must at lest partially wet its own solid.

{\em (ii) MD simulations:}
Choosing Al(111) as our test case, we have simulated its behavior at and above
$T_m$ using the recent accurate glue potential of Ercolessi and
Adams \cite{glueAl}, derived by fitting to first-principles data.
First, the approximate bulk melting point for this
potential was determined using the phase coexistence technique \cite{furio1}
and found to be $T_m=939\pm5\,{\rm K}$ (against an experimental value of
$T_m^{\rm exp}=933.6\,{\rm K}$).
Then a 16-layers slab with 3 rigid bottom layers, one free surface,
224 atoms per layer and $x$-$y$ periodic boundary conditions,
was studied as a function of $T$.
As $T_m$ was reached and crossed, the surface remained crystalline
(metastable) as expected, up to a large
$T_i=1088\pm18\,{\rm K}=T_m+ (149\pm18)\,{\rm K}$,
even for very long (2 ns) runs.
On the basis of our theory, using the known values of
$\rho=0.0534\,\rm\AA^{-3}$, $L=105.4\,\rm meV/atom$ \cite{glueAl}, and an
estimated $\xi=2.6\pm0.3\,\rm\AA$ \cite{orio}, we extract from (\ref{eq:ti})
$\Delta\gamma_\infty=2.3\pm0.4 \,\rm meV\AA^{-2}$.
Inserting in eq.\ (\ref{eq:thetam}), with an estimated value of
$\gamma_{\rm LV}=46.6\,\rm meV\AA^{-2}$ (obtained with a separate simulation
of the free liquid surface at $T=T_m$) we finally predict
$\theta_m=(18\pm2)^\circ$ and, with a value
$\gamma_{\rm SV}=54.3\,\rm meV\AA^{-2}$ \cite{glueAl},
$\theta_c=(16\pm2)^\circ$.

To check this prediction, we have prepared an 861 Al-atom cluster which
is fully melted and forms a liquid drop already at 900 K \cite{pawlow}.
By depositing this Al drop on any given Al surface, we can learn about
its wetting habit. We deposit it first on the Al(110) face, which is prone
to melting \cite{gonAl}. At $T=930\,\rm K$ (below $T_m$, but above
$T_w$), the drop spreads out completely within 100 ps
(fig.\ \ref{fig:cowshit}a-c).
However, when deposited on the nonmelting Al(111) face
it does not diffuse away, but rather
settles down as expected with well-defined exterior and interior angles
whose azimuthal average $\langle\theta\rangle$ we can extract.
By increasing temperature across $T_m$, from 930 to 945 K, we find that
$\langle\theta_{\rm LV}\rangle$ changes from $(24\pm3)^\circ$ to
$(21\pm1)^\circ$, and $\langle\theta_{\rm SV}\rangle$ from
$-\langle\theta_{\rm LV}\rangle$ (the droplet is essentially crystallized)
to $(44\pm6)^\circ$.
By interpolation we extract $\theta_m=(22\pm3)^\circ$, in fairly good agreement
with the predicted value $(18\pm2)^\circ$. The approximate values of $\xi$, and
of $\gamma_{\rm LV}$ (about 20\% lower than its experimental value, with our
potential) constitute sources of error.
Additional discrepancies are to be attributed to the macroscopic and
phenomenological nature of the theory, which should in principle be improved
to include fluctuations and finite-size effects.  On the simulation side,
one could consider in the future finite-size scaling as a possibility.

{\em (iii) Connection with experiments:}
We are not aware of measurements of $\theta_m$, $\theta_c$ or $T_i$
on Al(111), for which we have thus a direct prediction.
On Pb(111) van Pinxteren and Frenken \cite{helene} measured
$\theta_c=(14.7\pm1.4)^\circ$, whence using eq.\ (\ref{eq:thetac}),
and for $\gamma_{\rm SV}\simeq 34\,\rm meV\AA^{-2}$ we obtain
$\Delta \gamma_\infty=1.2\pm0.2\,\rm meV\AA^{-2}$.
Furthermore, using $\rho=0.033\,\rm\AA^{-3}$, $L=50\,\rm meV/atom$, and
$\xi\simeq 2.7\,\rm\AA$ (averaging data from ref. \cite{pluisXsi} as
suggested in \cite{helene}), we obtain
via eq.\ (\ref{eq:ti}) $T_i=T_m+(150\pm30)\,{\rm K}$.
This is in rather good agreement with the experimental result,
$T_i^{\rm exp} \simeq T_m^{\rm exp} + 120\,{\rm K}$ \cite{herman}. From
eq.\ (\ref{eq:thetam}), using $\gamma_{\rm LV}\simeq 28\,\rm meV\AA^{-2}$, we
also predict $\theta_m=(16\pm1)^\circ$ for a droplet on Pb(111) at $T=T_m$.
For Al(100), another nonmelting surface \cite{molen}, we find by simulation
$T_i=1025\pm5\,\rm K$, and assuming $\xi\simeq3\,\rm\AA$ we predict
$\theta_m\simeq15^\circ$ and $\theta_c\simeq13^\circ$.
For Cd(0001), where $\theta_m=(37\pm1)^\circ$ \cite{mut1},
and using $\rho=0.043\,\rm\AA^{-3}$,
$L=64\,\rm meV/atom$, $\gamma_{\rm LV}=40\,\rm meV\AA^{-2}$, and again a
guessed $\xi\simeq3\,\rm\AA$, we get $T_i\simeq T_m+580\,{\rm K}$,
close to twice the melting temperature (594 K).
Application of this scheme to Ge(111) or NaCl(111) appears instead
problematic, due to the essential role of long-range forces in these cases.
In fact Ge(111) has a negative Hamaker constant, which is probably
related to its nonmelting behavior \cite{take}, while long-range
Coulomb forces are likely to be relevant to the nonmelting of NaCl(100).
Finally, the present scheme is probably also inapplicable in its simplest form
to surfaces such as Pb(100) or Au(100) which undergo {\it incomplete melting}
\cite{pavlovska,helene100,ocko,bila100}.

In summary, we have described new results on the nonmelting crystal surfaces.
A simple theory of nonmelting is given, which describes the metastable
solid surface above $T_m$, up to a maximum overheating
temperature $T_i$ which acts as a spinodal point.
This temperature is found to have a simple connection with the partial
wetting angle of the surface by its own melt at $T\approx T_m$,
and with the tilting angle of melted regions in vicinals undergoing
nonmelting-induced faceting.
Computer simulations on Al(111), as well as available data on Pb(111) are in
good agreement with this theory.

We acknowledge instructive discussions with R.\ Evans, and support from EEC
through contracts ERBCHBGCT920180, ERBCHBGCT940636 and ERBCHRXCT930342,
from INFM, and from CNR under project SUPALTTEMP.

\clearpage

\begin{figure}
\caption{Critical liquid thickness of a nonmelting surface
vs. temperature above $T_m$ (schematic). A system with a liquid film thinner
than the critical value, will recrystallize for any $T$ between
$T_m$ and $T_i$. One with a thicker film will melt completely.
Inset: free energy change upon conversion of a film  of thickness $\ell$
from solid to liquid. From $T_m$ to $T_i$ the solid surface is a local
minimum.}
\label{fig:df}
\end{figure}

\begin{figure}
\caption{Shape of a drop of melt onto a nonmelting solid surface of
the same material. The two interfaces separating solid and liquid (SL), and
liquid and vapor (LV) are assumed to be spherical, with radii
$R_{\rm SL}$ and $R_{\rm LV}$ and contact angles
$\theta_{\rm SL}$ and $\theta_{\rm LV}$ respectively.}
\label{fig:drop}
\end{figure}

\begin{figure}
\caption{Evolution of an 861-particles liquid drop of Al on substrate of the
same material.
Left column: drop on a surface undergoing surface melting (Al(110) at
$T=0.99\, T_m$). (a) before contact; (b) after contact, the drop spreads
readily; (c) the drop has been fully absorbed.
Right column: drop on a nonmelting overheated surface (Al(111) at
$T=1.01\, T_m$). (d) before contact; (e) after contact: the drop settles,
but does not spread; (f) final drop shape.
Darkness of atoms is proportional to their square displacement in the run.}
\label{fig:cowshit}
\end{figure}

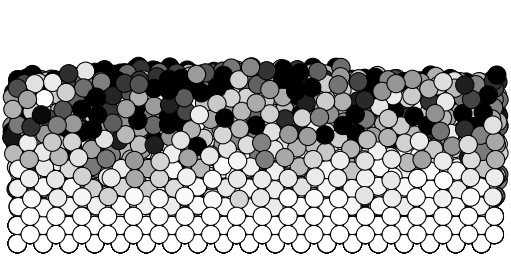
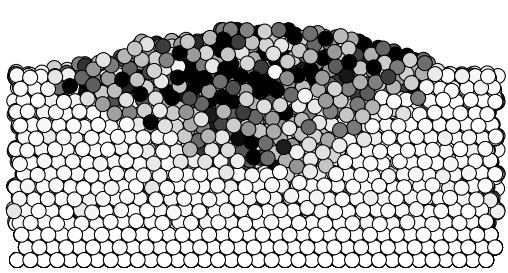
\end{document}